\documentclass[review]{elsarticle}

\usepackage{lineno,hyperref}
\modulolinenumbers[5]

\usepackage{amsmath,amsthm,amssymb}
\usepackage{graphics,color}

\renewcommand{\vec}[1]{\boldsymbol{#1}}

\newcommand{\ket}[1]{|#1\rangle}
\newcommand{\braket}[2]{\langle#1|#2\rangle}

\newcommand{\mtx}[1]{\boldsymbol{\mathsf{#1}}}

\journal{Physics Letters A}

\begin{document}

\begin{frontmatter}

\title{Local Hidden-Variable Model for a Recent Experimental Test of Quantum Nonlocality and Local Contextuality}

\author{Brian R. La Cour}
\address{Applied Research Laboratories, 
               The University of Texas at Austin, 
               P.O. Box 8029, 
               Austin, TX 78713-8029}
\fntext[myfootnote]{blacour@arlut.utexas.edu}

\begin{abstract}
\textcolor{black}{An experiment has recently been performed to demonstrate quantum nonlocality by establishing contextuality in one of a pair of photons encoding four qubits}; however, low detection efficiencies and use of the fair-sampling hypothesis leave these results open to possible criticism due to the detection loophole.  In this Letter, a physically motivated local hidden-variable model is considered as a possible mechanism for explaining the experimentally observed results.  The model, though not intrinsically contextual, acquires this quality upon post-selection of coincident detections.
\end{abstract}

\begin{keyword}
quantum nonlocality \sep contextuality \sep entanglement \sep quantum measurement
\end{keyword}

\end{frontmatter}


\section{Introduction}

Quantum nonlocality and quantum contextuality are intimately related in a manner that had not been appreciated for some time.  Both are used in the construction of various no-go theorems for ruling out different classes of hidden variable models, yet the two properties are, in some sense, very much intertwined.  The Bell-Kochen-Specker theorem demonstrates that quantum mechanics is fundamentally contextual in the sense that it is inconsistent with a hidden-variable model that does not exhibit contextuality \cite{Bell1966,KochenSpecker1967,Mermin1993}.  These so-called noncontextual hidden-variable (NCHV) models may be characterized as having a probability distribution over the hidden variable space that is independent of the choice of measurement basis.  In a similar manner, the Bell inequality is obeyed so long as the probability distribution over the hidden variable space is the same for all choices of measurement settings \cite{Bell1964}.  Thus, violations of the Bell inequality may be seen as a signature of contextuality \cite{Cabello2008}.  If, furthermore, the invariance of the probability distribution can be justified on the grounds of local realism, then such violations may be seen as a signature of nonlocality, meaning that they are inconsistent with any local hidden-variable (LHV) theory \cite{Brunner2014}.

The difficulty with these no-go theorems is that contextuality can arise in subtle ways that may have nothing to do with quantum mechanics.  One of the best examples of this comes from post-selection.  In experiments using entangled photons, one often post-selects on outcomes for which a coincident detection of both photons is achieved \cite{CHSH1969,AGR1982,Weihs1998,Scheidl2010,Giustina2013}.  Doing so, however, creates a situation in which, from a hidden-variable perspective, different subensembles are downselected for each measurement setting.  If one then adopts the fair-sampling hypothesis, then one is asserting, without independent justification, that these subensembles are in fact the same and, hence, that there is no contextuality.  A subsequent violation of a Bell inequality, then, leaves open the question of whether the assumption of noncontextuality was indeed correct.  This, of course, is the origin of the so-called detection loophole \cite{Pearle1970,Garg&Mermin1987,Larsson1998a}.  Although the detection loophole has been closed in some experiments \cite{Christensen2013,Hensen2015,Giustina2015,Shalm2015}, this is not true for many cases and, so, the matter of their interpretation is left open.

Recently, the connection between nonlocality and contextuality was studied experimentally using pairs of photons that were exquisitely prepared in a \textcolor{black}{four-qubit} state involving both polarization and spatial modes \cite{Liu2016}.  The resulting quantum state may be thought of as a four-qubit system, with the first two qubits corresponding to the polarization and spatial modes of one photon and the second two qubits corresponding to those of the other photon.  Using an experimental design developed by Cabello \cite{Cabello2010}, the two photons were then each subjected to a set of two-qubit measurements chosen so that their outcomes would satisfy certain Bell-like inequalities whose violation would be indicative of contextuality.  The authors conclude that ``there are correlations in nature which cannot be explained by LHV theories \emph{because} they contain single-particle correlations which cannot be reproduced with NCHV theories'' \cite{Liu2016}.

This conclusion appears premature given the experimenters' reliance on the fair sampling assumption.  An interesting and still open question is whether the fair sampling assumption is, indeed, valid.  This question is of general physical interest and quite independent of whether one has closed the detection loophole or not.  Given the apparent reasonableness of this assumption, an investigation of specific LHV models may help to shed some light on whether it is, indeed, reasonable to suppose that detected photons are statistically identical to their undetected kin.

In this Letter, a previously described LHV model is used to reproduce the results of this experiment under similar experimental conditions \cite{LaCour2014}.  This is made possible by virtue of the fact that, like the experimenters, we restrict consideration to coincident detections only, thus giving rise to contextuality as an emergent property of the post-selection process.  Variations of this LHV model are described elsewhere and have been used to explain the appearance of contextuality and nonlocality in entangled photon experiments \cite{LaCour2016}.


\section{Description of the Experiment}

The experiment of Ref.\ \cite{Liu2016} may be described in terms of a four-qubit system.  Consider a sequence of four bits $x_1, x_2, x_3, x_4 \in \{0,1\}$ used to index one of 16 basis states, each of which is written as
\begin{equation}
\ket{x_1}_1 \otimes \ket{x_2}_2 \otimes \ket{x_3}_3 \otimes \ket{x_4}_4 = \ket{x_1 x_2 x_3 x_4} \; .
\end{equation}
In the context of the experiment, qubits 1 and 2 correspond to the polarization and spatial modes, respectively, of the qubit measured by Alice, while qubits 3 and 4 correspond to those measured by Bob.  A hyperentangled state is prepared that may be described as follows:
\begin{equation}
\ket{\Psi_{1234}} = \frac{1}{2} \Bigl[ \ket{0011} - \ket{0110} - \ket{1001} + \ket{1100} \Bigr] \; .
\label{eqn:Psi}
\end{equation}

The experiment now considers combinations of the following nine two-qubit observables:
\begin{subequations}
\begin{eqnarray}
A = \mtx{Z} \otimes \mtx{I}_2 \quad &B = \mtx{I}_2 \otimes \mtx{Z} \quad &C = \mtx{Z} \otimes \mtx{Z} \\
a = \mtx{I}_2 \otimes \mtx{X} \quad &b = \mtx{X} \otimes \mtx{I}_2 \quad &c = \mtx{X} \otimes \mtx{X} \\
\alpha = \mtx{Z} \otimes \mtx{X} \quad &\beta = \mtx{X} \otimes \mtx{Z} \quad &\gamma = \mtx{Y} \otimes \mtx{Y}
\end{eqnarray}
\label{eqn:magic}
\end{subequations}
where $\mtx{I}_2, \mtx{X}, \mtx{Y}, \mtx{Z}$ are the Pauli spin matrices and $\otimes$ denotes the Kronecker product between matrices.  These observables form a Mermin-Peres magic square such that $AB = C$, $ab = c$, $Aa = \alpha$, $Bb = \beta$, but $\alpha \beta = \gamma = -Cc$.  Each of the six rows and columns comprises a compatible set of observables and, hence, may be measured in a common basis.  Such constructions have been used extensively to study quantum contextuality \cite{Mermin1990,Peres1990}.

A measurement of $A$ by Alice, who has local access only to qubits 1 and 2, is denoted by the $16\times16$ matrix $A \otimes \mtx{I}_4$, where $\mtx{I}_4$ is the $4\times4$ identity matrix.  Similarly, a measurement of $A$ by Bob, who has local access only to qubits 3 and 4, is denoted by $\mtx{I}_4 \otimes A$.  Thus, a measurement of $A$ by Alice and $A$ by Bob corresponds to the separable observable $(A \otimes \mtx{I}_4)(\mtx{I}_4 \otimes A) = A \otimes A$.  (In Ref.\ \cite{Liu2016}, $A \otimes \mtx{I}_4$ is denoted $A$, and $\mtx{I}_4 \otimes A$ is denoted $A'$, so the product of the two is there denoted $AA'$.  Here we write the Kronecker product explicitly for clarity.)  Each of these measurements can be performed in one of two experimental contexts, corresponding to the intersecting row or column in the magic square.

In the experiment, Alice measures all three observables in the chosen basis.  Bob, however, measures only one, so his choice of basis is irrelevant.  Following the notation of Ref.\ \cite{Liu2016}, the choice of basis for Alice will be denoted by one of $CAB, cba, \beta\gamma\alpha$ for the three rows and $\alpha Aa, \beta bB, c\gamma C$ for the three columns.  Thus, six different sets of measurements are performed, each corresponding to one of the six basis choices for Alice.  From these, two averaged quantities are measured, $\langle \chi \rangle$ and $\langle S \rangle$.  These are combined to form a single quantity, $\langle \omega \rangle = \langle \chi \rangle + \langle S \rangle$, which, according to Cabello, satisifies the inequality $\langle \omega \rangle \le 16$ for any LHV model \cite{Cabello2010}.

The quantity $\langle \chi \rangle$ is given solely in terms of Alice's observables and is defined as
\begin{multline}
\langle \chi \rangle = \langle CAB \otimes \mtx{I}_4 \rangle + \langle cba \otimes \mtx{I}_4 \rangle + \langle \beta\gamma\alpha \otimes \mtx{I}_4 \rangle \\
+ \langle {\alpha}Aa \otimes \mtx{I}_4 \rangle + \langle \beta bB \otimes \mtx{I}_4 \rangle - \langle c\gamma C \otimes \mtx{I}_4 \rangle \; .
\end{multline}
For any quantum state, the ideal quantum predictions for the first five terms are each $+1$, while that for the last is $-1$, thereby yielding a maximal value of $\langle \chi \rangle = 6$.  According to Cabello, if the measured system exhibits no contextuality then the inequality $\langle \chi \rangle \le 4$ must hold \cite{Cabello2008}.  Thus, any observed violation of this latter inequality is an indication of contextuality.  The experimentally measured value for $\langle \chi \rangle$ was $5.817 \pm 0.011$, thus showing a clear violation of this inequality.

The quantity $\langle S \rangle$ is given in terms of observables for both Alice and Bob and is defined as
\begin{equation}
\begin{split}
\langle S \rangle &= -\langle A \otimes A \rangle_{CAB} - \langle B \otimes B \rangle_{CAB} \\
&\quad - \langle b \otimes b \rangle_{cba} - \langle a \otimes a \rangle_{cba} \\
&\quad + \langle \gamma \otimes \gamma \rangle_{\beta\gamma\alpha} + \langle \alpha \otimes \alpha \rangle_{\beta\gamma\alpha} \\
&\quad - \langle A \otimes A \rangle_{\alpha Aa} - \langle a \otimes a \rangle_{\alpha Aa} \\
&\quad - \langle b \otimes b \rangle_{\beta bB} - \langle B \otimes B \rangle_{\beta bB} \\
&\quad + \langle \gamma \otimes \gamma \rangle_{c\gamma C} + \langle C \otimes C \rangle_{c\gamma C} \; .
\end{split}
\end{equation}
Note that the subscripts on each expectation value are simply a reminder of the measurement context; in truth, each uses the same quantum state $\ket{\Psi}$ given by Eqn.\ (\ref{eqn:Psi}).  For this state, the ideal quantum predictions for the twelve terms are $\langle A \otimes A \rangle_{CAB} = \langle A \otimes A \rangle_{\alpha A a} = -1$, $\langle B \otimes B \rangle_{CAB} = \langle B \otimes B \rangle_{\beta bB} = -1$, $\langle a \otimes a \rangle_{cba} = \langle a \otimes a \rangle_{\alpha Aa} = -1$, $\langle b \otimes b \rangle_{cba} = \langle b \otimes b \rangle_{\beta bB} = -1$, $\langle \gamma \otimes \gamma \rangle_{\beta\gamma\alpha} = 
\langle \gamma \otimes \gamma \rangle_{c\gamma C} = +1$, $\langle \alpha \otimes \alpha \rangle_{\beta\gamma\alpha} = +1$, and $\langle C \otimes C \rangle_{c\gamma C} = +1$, yielding a maximal value of $\langle S \rangle = 12$.  The experimentally measured value for $\langle S \rangle$ was $11.430 \pm 0.016$.  Combined with the result for $\langle \chi \rangle$, this gives a value for $\langle \omega \rangle$ of $17.247 \pm 0.019$, in clear violation of the aforementioned inequality and, therefore, interpreted as a signature of quantum nonlocality.

These results, while statistically significant, where obtained under experimental conditions such that the overall detection efficiency was found to be only 3.3\%.  As the authors acknowledge, such low detection efficiency, combined with the fair sampling assumption, opens up the detection loophole.  They do note, however, that replacing the avalanche photodiode detectors used in the experiment with superconducting detectors, which can have efficiencies of over 95\%, should suffice to close this loophole.


\section{LHV Model}

For our LHV model, let $\vec{\lambda} \in \Lambda \subset \mathbb{C}^{16}$ be a $16\times1$ complex vector denoting the hidden variable state, each element of which may be indexed by a four-bit integer string $x_1x_2x_3x_4$ and defined such that
\begin{equation}
\lambda_{x_1x_2x_3x_4} = s (\sqrt{2}-1) \braket{x_1x_2x_3x_4}{\Psi_{1234}} + \nu_{x_1x_2x_3x_4} \; ,
\end{equation}
where $s > 0$ is a model tuning parameter and $\vec{\nu}$ is a normalized complex standard Gaussian random vector.  The factor of $\sqrt{2}-1$ ensures that, for $s \le 1$, we have $\|\vec{\lambda}\|^2 \le 2$, since $\|\vec{\nu}\| \le 1$.  We shall denote by $\Pr[\cdot]$ the resulting probability distribution of the hidden variables.  

The physical motivation for this model is as follows.  We imagine the photons as being described by classical electromagnetic waves, much as is done in the field of stochastic electrodynamics \cite{EmergingQuantum}.  The first term in $\vec{\lambda}$ represents the user-defined portion of the waves.  The random term $\vec{\nu}$ is intended to represent an uncontrolled contribution to these waves that is introduced during the state preparation process.  The origin of this random term is imagined to arise from the various modes of a zero-point field and transformed through the optical devices used for state preparation.  It should be emphasized that this is only a notional picture; the LHV model itself may simply be taken as a mathematical artifice.

Measurements are defined in terms of threshold exceedances for projection operators.  Thus, for a $16\times16$ projection matrix $\mtx{P}$, we say that an outcome of 1 is obtained for the hidden variable state $\vec{\lambda}$ if $\vec{\lambda}^\dagger \mtx{P} \vec{\lambda} > 1$.  If, however, $\vec{\lambda}^\dagger \mtx{P} \vec{\lambda} \le 1$, this does not necessarily imply an outcome of 0, as non-detection events are also possible.  To define the 0 outcome we consider the complementary projection matrix $\mtx{Q}$, defined such that $\mtx{P} + \mtx{Q}$ is the identity.  Now, if $\vec{\lambda}^\dagger \mtx{Q} \vec{\lambda} > 1$ we say that an outcome of 0 is obtained.  If neither $\vec{\lambda}^\dagger \mtx{P} \vec{\lambda}$ nor $\vec{\lambda}^\dagger \mtx{Q} \vec{\lambda}$ exceeds unity, we say that a non-detection has occurred.  If \emph{both} exceed unity, we say that a double-detection event has occurred.  (This might be interpreted as a multi-photon event.)  This is only possible when $s > 1$, which follows as a corollary of the following theorem.

\vspace{1em}
\noindent\textit{Theorem 1} If $\vec{\lambda}^\dagger \mtx{P} \vec{\lambda} > 1$ and $s \le 1$, then $\vec{\lambda}^\dagger \mtx{P}_\bot \vec{\lambda} \le 1$ for every $\mtx{P}_\bot$ orthogonal to $\mtx{P}$.

\textit{Proof} Suppose $\vec{\lambda}^\dagger \mtx{P} \vec{\lambda} > 1$ and $\vec{\lambda}^\dagger \mtx{P}_\bot \vec{\lambda} > 1$ for some $\mtx{P}_\bot$ such that $\mtx{P}_\bot \mtx{P} = \mtx{0}$.  Since $\mtx{P}$ and $\mtx{P}_\bot$ are orthogonal projections, $\|\vec{\lambda}\|^2 \ge \| \mtx{P} \vec{\lambda} + \mtx{P}_\bot \vec{\lambda} \|^2 = \vec{\lambda}^\dagger \mtx{P}^2 \vec{\lambda} + \vec{\lambda}^\dagger \mtx{P}_\bot^2 \vec{\lambda} + \vec{\lambda}^\dagger \mtx{P}\mtx{P}_\bot \vec{\lambda} + \vec{\lambda}^\dagger \mtx{P}_\bot\mtx{P} \vec{\lambda} = \vec{\lambda}^\dagger \mtx{P} \vec{\lambda} + \vec{\lambda}^\dagger \mtx{P}_\bot \vec{\lambda} >  2$.  But $\| \vec{\lambda} \|^2 \le 2$ for $s \le 1$.  This proves the theorem.
\vspace{1em}

Thus, if we have several projection matrices $\mtx{P}_1, \mtx{P}_2, \ldots$ that are all mutually orthogonal, then, for $s \le 1$, at most one of these will have a threshold exceedance.  This property mimics the particle-like behavior of a single-photon, multi-mode system.


\section{Application of LHV Model}

In the LHV model, both Alice and Bob are assumed to have access to the entire hidden variable $\vec{\lambda}$ of a particular realization.  Locality is enforced by insisting that Alice can only perform operations on ``her'' qubits (1 and 2), while Bob can only perform operations on ``his'' qubits (3 and 4).

For example, to measure the last term of $\langle \chi \rangle$, Alice must measure the observable $c \gamma C$.  She does this by first constructing the following eight sets:
\begin{align*}
\tilde{I}^{+++}_{c\gamma C} &= \{ \vec{\lambda} \in \Lambda : \vec{\lambda}^\dagger (\mtx{P}_c \mtx{P}_\gamma \mtx{P}_C \otimes \mtx{I}_4) \vec{\lambda} > 1 \} \\
\tilde{I}^{++-}_{c\gamma C} &= \{ \vec{\lambda} \in \Lambda : \vec{\lambda}^\dagger (\mtx{P}_c \mtx{P}_\gamma \mtx{Q}_C \otimes \mtx{I}_4) \vec{\lambda} > 1 \} \\
&\vdots \\
\tilde{I}^{---}_{c\gamma C} &= \{ \vec{\lambda} \in \Lambda : \vec{\lambda}^\dagger (\mtx{Q}_c \mtx{Q}_\gamma \mtx{Q}_C \otimes \mtx{I}_4) \vec{\lambda} > 1 \} \; .
\end{align*}
For $s \le 1$, these sets are disjoint, as shown in Theorem 1.  For higher values of $s$, ``multi-photon'' detections (in which two or more of the above sets overlap) may be possible.  In our post-selection analysis, we will discard such instances by replacing the above sets with the following:
\begin{equation}
I^{\sigma_1 \sigma_2 \sigma_3}_{c \gamma C} = \tilde{I}^{\sigma_1 \sigma_2 \sigma_3}_{c\gamma C} \; \setminus \bigcup_{\sigma_1' \sigma_2' \sigma_3' \neq \sigma_1 \sigma_2 \sigma_3} \tilde{I}^{\sigma_1' \sigma_2' \sigma_3'}_{c \gamma C} \; .
\end{equation}
So, $\vec{\lambda} \in I^{++-}_{c\gamma C}$, say, means Alice obtains a single detection with outcomes of $+1$ for $c$, $+1$ for $\gamma$, and $-1$ for $C$.

Bob similarly performs a measurement of, say, $\gamma$ on qubits 3 and 4 by constructing the sets
\begin{subequations}
\begin{align}
\tilde{J}^{+}_{\gamma} &= \{ \vec{\lambda} \in \Lambda : \vec{\lambda}^\dagger (\mtx{I}_4 \otimes \mtx{P}_\gamma) \vec{\lambda} > 1 \} \; , \\
\tilde{J}^{-}_{\gamma} &= \{ \vec{\lambda} \in \Lambda : \vec{\lambda}^\dagger (\mtx{I}_4 \otimes \mtx{Q}_\gamma) \vec{\lambda} > 1 \} \; .
\end{align}
\end{subequations}
We again post-select out any multiple detections by using
\begin{equation}
J^+_\gamma = \tilde{J}^+_\gamma \setminus \tilde{J}^-_\gamma \quad \text{and} \quad J^-_\gamma = \tilde{J}^-_\gamma \setminus \tilde{J}^+_\gamma \; .
\end{equation}
So, $\vec{\lambda} \in J^+_{\gamma}$ means that Bob obtains a single detection with an outcome of $+1$.  In this example, we do not actually care which outcome Bob obtains, so long as he detects something, so we will instead consider the union of these two sets.

We are interested in when the product of Alice's three outcomes is either $+1$ or $-1$.  Thus, we consider the sets
\begin{subequations}
\begin{align}
I^+_{c \gamma C} &= I^{+++}_{c \gamma C} \, \cup \, I^{+--}_{c \gamma C} \, \cup \, I^{-+-}_{c \gamma C} \, \cup \, I^{--+}_{c \gamma C} \; , \\
I^-_{c \gamma C} &= I^{---}_{c \gamma C} \, \cup \, I^{-++}_{c \gamma C} \, \cup \, I^{+-+}_{c \gamma C} \, \cup \, I^{++-}_{c \gamma C} \; .
\end{align}
\end{subequations}
Now, the event that Alice gets a single detection with a product of $\pm1$ and Bob gets a single detection is
\begin{equation}
K^\pm_{c \gamma C, \gamma} = I^\pm_{c \gamma C} \, \cap \, ( J^+_\gamma \cup J^-_\gamma ) \; .
\end{equation}
Note that this event can only be determined once Alice and Bob have classically communicated their individual results as part of the post-selection process.

The expectation of the random variable $v(c\gamma C \otimes \mtx{I}_4): \Lambda \mapsto \mathbb{R}$ corresponding to the operator $c\gamma C \otimes \mtx{I}_4$, when conditioned on single, coincident detections, can now be computed as follows:
\begin{equation}
E_{c\gamma C,\gamma}\bigl[ v(c\gamma C \otimes \mtx{I}_4) \bigr] = \frac{\Pr\bigl[ K^+_{c \gamma C, \gamma} \bigr] - \Pr\bigl[ K^-_{c \gamma C, \gamma} \bigr]}{\Pr\bigl[ K^+_{c \gamma C, \gamma} \cup K^-_{c \gamma C, \gamma} \bigr]}
\end{equation}
Now, since $c\gamma C = -\mtx{I}_4$, it follows that
\begin{equation}
\mtx{P}_c \mtx{P}_\gamma \mtx{P}_C + \mtx{P}_c \mtx{Q}_\gamma \mtx{Q}_C + \mtx{Q}_c \mtx{P}_\gamma \mtx{Q}_C + \mtx{Q}_c \mtx{Q}_\gamma \mtx{P}_C = \mtx{0}_{12} \; .
\end{equation}
while
\begin{equation}
\mtx{Q}_c \mtx{Q}_\gamma \mtx{Q}_C + \mtx{Q}_c \mtx{P}_\gamma \mtx{P}_C + \mtx{P}_c \mtx{Q}_\gamma \mtx{P}_C + \mtx{P}_c \mtx{P}_\gamma \mtx{Q}_C = \mtx{I}_4 \; ,
\end{equation}
Hence, $I^+_{c \gamma C} = \varnothing$, so $E_{c\gamma C,\gamma}\bigl[ v(c\gamma C \otimes \mtx{I}_4) \bigr] = -1$.
Note that this result is independent of which observable Bob chooses to measure.  It is also independent of both $s$ and the initial quantum state.  The expectation values of the other five observables for $\langle \chi \rangle$ are found similarly and give $+1$ each.  Thus, the sum of the first five terms, subtracted by the last term, is $6$.  This matches precisely the ideal quantum prediction.

The terms in $\langle S \rangle$ are determined in a similar manner.  For example, the second-to-last term gives the expectation of the observable $\gamma \otimes \gamma = (\mtx{P}_\gamma - \mtx{Q}_\gamma) \otimes (\mtx{P}_\gamma - \mtx{Q}_\gamma)$ when Alice measures $c \gamma C$.  In the LHV model, this is estimated as follows.  First, Alice computes the sets
\begin{align}
I^+_{c \gamma C,\gamma} &= I^{+++}_{c \gamma C} \cup I^{++-}_{c \gamma C} \cup I^{-++}_{c \gamma C} \cup I^{-+-}_{c \gamma C} \; ,\\
I^-_{c \gamma C,\gamma} &= I^{+-+}_{c \gamma C} \cup I^{+--}_{c \gamma C} \cup I^{--+}_{c \gamma C} \cup I^{---}_{c \gamma C} \; ,
\end{align}
where $I^{\pm}_{c \gamma C,\gamma}$ corresponds to an outcome of $\pm1$ for $\gamma$ when Alice measures $c$, $\gamma$, and $C$.  Note that these sets are different from the sets $I^{\pm}_{c \gamma C}$ corresponding to the product of the three observables.  So, while the set $I^{+}_{c \gamma C}$ is empty, $I^{+}_{c \gamma C,\gamma}$ need not be.

Next, Alice and Bob compare their results and compute the following sets:
\begin{subequations}
\begin{align}
L^+_{c \gamma C,\gamma} &= ( I^+_{c \gamma C,\gamma} \cap J^+_\gamma ) \, \cup ( I^-_{c \gamma C} \cap J^-_\gamma ) \; , \\
L^-_{c \gamma C,\gamma} &= ( I^+_{c \gamma C,\gamma} \cap J^-_\gamma ) \, \cup ( I^-_{c \gamma C} \cap J^+_\gamma ) \; .
\end{align}
\end{subequations}
Thus, $L^+_{c \gamma C,\gamma}$, say, is the event that Alice and Bob get single detections with the same outcome (either both $+1$ or both $-1$).  Again, this can only be determined  through the post-selection process.  The expectation of the random variable $v(\gamma \otimes \gamma)$, conditioned on single, coincident detections for Alice and Bob, is now
\begin{equation}
E_{c \gamma C,\gamma}[v(\gamma \otimes \gamma)] = \frac{\Pr\bigl[ L^+_{c \gamma C, \gamma} \bigr] - \Pr\bigl[ L^-_{c \gamma C, \gamma} \bigr]}{\Pr\bigl[ L^+_{c \gamma C, \gamma} \cup L^-_{c \gamma C, \gamma} \bigr]} \; .
\end{equation}

By contrast, the expectation value of $v(\gamma \otimes \gamma)$ conditioned on single, coincident detections when Alice measures $\beta, \gamma, \alpha$ and Bob measures $\gamma$ is now given by
\begin{equation}
E_{\beta \gamma \alpha,\gamma}[v(\gamma \otimes \gamma)] = \frac{\Pr\bigl[ L^+_{\beta \gamma \alpha, \gamma} \bigr] - \Pr\bigl[ L^-_{\beta \gamma \alpha, \gamma} \bigr]}{\Pr\bigl[ L^+_{\beta \gamma \alpha, \gamma} \cup L^-_{\beta \gamma \alpha, \gamma} \bigr]} \; .
\end{equation}
These two expectation values differ only in the manner in which the post-selection is performed, which results in a potentially different conditional distribution.  At the time of measurement, Alice need not communicate to Bob whether she is measuring $c, \gamma, C$ or $\beta, \gamma, \alpha$, but the post-selected results will reflect this choice.  The other terms in $\langle S \rangle$ are computed in a similar manner.

Detection efficiency, $\eta$, is defined as the ratio of coincident detections to Bob's single detections, minimized over all measurement contexts.  For a specific measurement context, we have
\begin{equation}
\eta_{c \gamma C} = \frac{\Pr[(I^+_{c \gamma C} \cup I^-_{c \gamma C}) \cap (J^+_\gamma \cup J^-_\gamma)]}{\Pr[J^+_\gamma \cup J^-_\gamma]} \; .
\end{equation}
The overall detection efficiency is therefore given by
\begin{equation}
\eta = \min\left\{ \eta_{CAB}, \, \eta_{cba}, \, \eta_{\beta\gamma\alpha}, \, \eta_{\alpha A a}, \, \eta_{\beta b B}, \, \eta_{c \gamma C} \right\} \; .
\end{equation}

Similarly, we may define the multi-photon detection probability, $\epsilon$, as the ratio of multi-photon coincident detections from Alice over Bob's single detections, maximized over all measurement contexts.  For a specific measurement context, we have
\begin{equation}
\epsilon_{c \gamma C} = \Pr\left[ (J^+_\gamma \cup J^-_\gamma)  \cap \bigcup_{\sigma_1\sigma2\sigma_3} \tilde{I}^{\sigma_1\sigma_2\sigma_3}_{c \gamma C} \setminus (I^+_{c \gamma C} \cup I^-_{c \gamma C}) \right] / \Pr[J^+_\gamma \cup J^-_\gamma] \; .
\end{equation}
Thus, the overall multi-photon detection probability is
\begin{equation}
\epsilon = \max\left\{ \epsilon_{CAB}, \, \epsilon_{cba}, \, \epsilon_{\beta\gamma\alpha}, \, \epsilon_{\alpha A a}, \, \epsilon_{\beta b B}, \, \epsilon_{c \gamma C} \right\} \; .
\end{equation}

For the Alice-Bob correlations, analytic results are not readily available.  Instead, a numerical simulation was performed to estimate the expectation values as a function of the scaling parameter $s$ \cite{SuppMat}.  In the simulation, a set of $N = 2^{20} \sim 10^6$ independent realizations of the random vector $\vec{\lambda}$ were drawn.  This ensemble was kept fixed and was therefore independent of which measurements were to be performed, although the results are qualitatively the same if different realizations are used.

\begin{figure}[ht]
\centerline{\scalebox{0.6}{\includegraphics{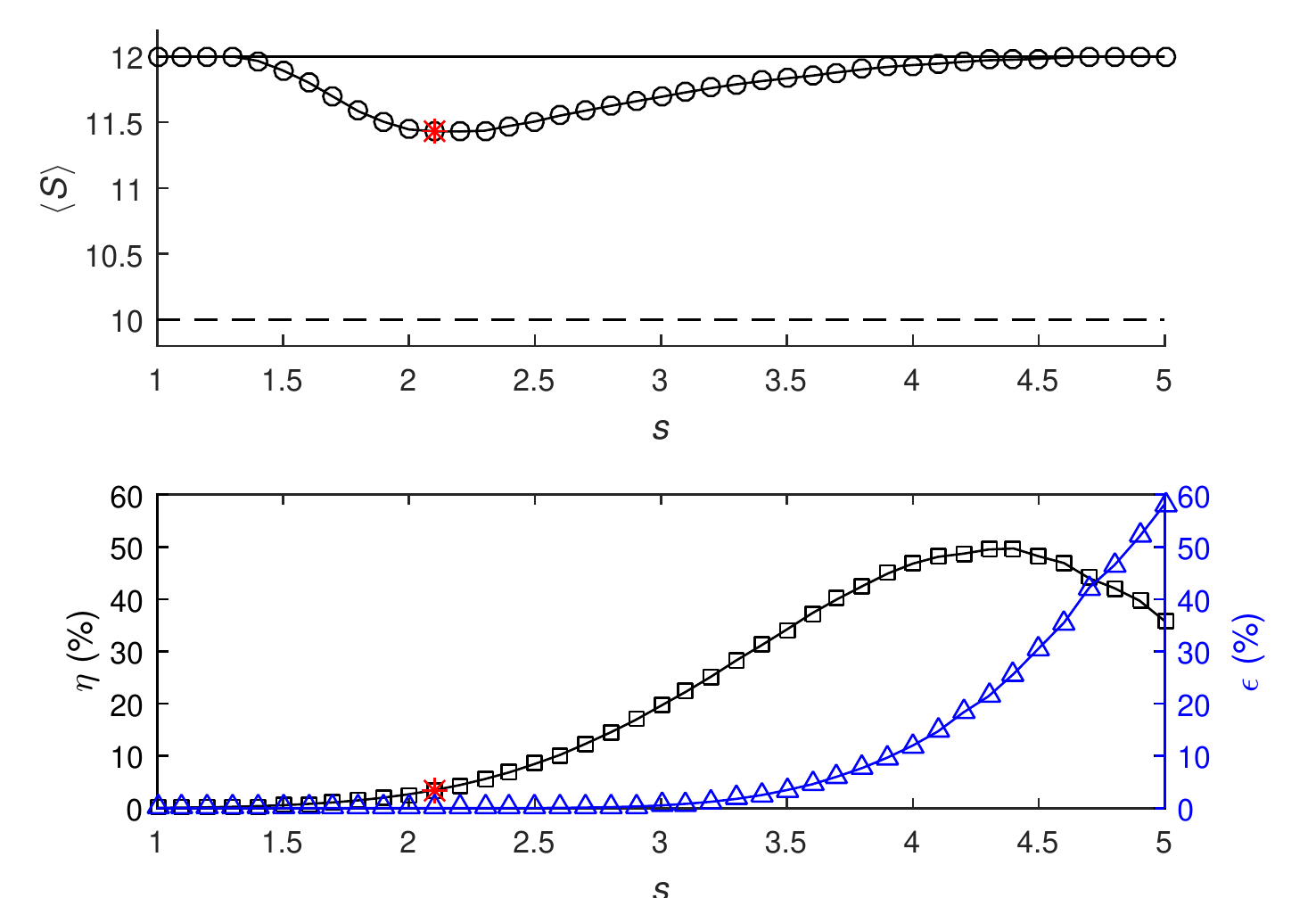}}}
\caption{(color online) Plot of $\langle S \rangle$ (upper figure) and $\eta$, $\epsilon$ (lower figure) for the LHV model versus the scale factor $s$.  The solid and dashed horizontal lines in the upper figure are the maximum value (12) and contextuality threshold (10), respectively.  The black squares and blue triangles in the lower figure indicate the detection efficiency ($\eta$) and multi-photon probability ($\epsilon$), respectively.  The red asterisks indicate experimental results from Ref.\ \cite{Liu2016}, fitted to a value of $s = 2.1$.}
\label{fig:figure1}
\end{figure}

The results are summarized in Fig.\ \ref{fig:figure1}, where we have plotted the value of $\langle S \rangle$, as estimated by the LHV model, as well as the detection efficiency, $\eta$, as a function of the fitting parameter $s$.  Also plotted is the multi-photon detection probability, $\epsilon$.  For small values of $s$ (less than or equal to about 1.4), a nearly maximal value of 12 is obtained, though with a detection efficiency of less that 0.5\%.  For the intermediate value of $s = 2.1$, we find good agreement with the experimental results of Ref.\ \cite{Liu2016} for both $\langle S \rangle$, reported at 11.430, and the efficiency, reported to be 3.3\%.  These values are shown in Fig.\ \ref{fig:figure1} as well for comparison.  Still larger values of $s$ show an increase in both $\langle S \rangle$ and $\eta$, with both peaking at around $s = 4.5$.  Thereafter, the number of single-photon detections begins to fall off.  In all cases, however, $\langle S \rangle$ remains well above the noncontextuality threshold.


\section{Conclusion}

This Letter has described a local hidden variable model capable of simultaneously exhibiting contextuality and violating certain Bell-like inequalities.   The key enabling factor was the use of post-selection on single-photon coincident detections, as was done in this and many other single-photon experiments.  Furthermore, with a single fitting parameter we found surprisingly good agreement with both the reported violation and the observed detection efficiency of Ref.\ \cite{Liu2016}.  Although ideal, maximal violations are possible with this model, the maximum detection efficiency that can be achieved was found to be less than 50\%.  Thus, experiments showing a violation of the noncontextuality bound with a larger coincident detection efficiency would be inconsistent with the present model.


\section*{Acknowledgments}

This work was support by the Office of Naval Research under Grant Nos.\ N00014-14-1-0323 and N00014-17-1-2107.  The author would like to thank A.\ Cabello for his insightful comments and helpful suggestions.


\section*{References}


\begin{thebibliography}{10}
\expandafter\ifx\csname url\endcsname\relax
  \def\url#1{\texttt{#1}}\fi
\expandafter\ifx\csname urlprefix\endcsname\relax\def\urlprefix{URL }\fi
\expandafter\ifx\csname href\endcsname\relax
  \def\href#1#2{#2} \def\path#1{#1}\fi

\bibitem{Bell1966}
J.~S. Bell, On the problem of hidden variables in quantum mechanics, Rev.\
  Mod.\ Phys. \textbf{38} (1966) 447.

\bibitem{KochenSpecker1967}
S.~Kochen, E.~Specker, The problem of hidden variables in quantum mechanics, J.
  Math.\ Mech. \textbf{17} (1967) 59.

\bibitem{Mermin1993}
N.~D. Mermin, Hidden variables and the two theorems of {J}ohn {B}ell, Rev.\
  Mod.\ Phys. \textbf{65} (1993) 803.

\bibitem{Bell1964}
J.~S. Bell, On the {E}instein-{P}odolsky-{R}osen paradox, Physics \textbf{1}
  (1964) 195.

\bibitem{Cabello2008}
A.~Cabello, Experimentally testable state-independent quantum contextuality,
  Phys.\ Rev.\ Lett. \textbf{101} (2008) 210401.

\bibitem{Brunner2014}
N.~Brunner, D.~Cavalcanti, S.~Pironio, V.~Scarani, S.~Wehner, Bell nonlocality,
  Rev.\ Mod.\ Phys. \textbf{86} (2014) 419.

\bibitem{CHSH1969}
J.~F. Clauser, M.~A. Horne, A.~Shimony, R.~A. Holt, Proposed experiment to test
  local hidden-variable theories, Phys. Rev. Lett. \textbf{23} (1969) 880.

\bibitem{AGR1982}
A.~Aspect, P.~Grangier, G.~Roger, Experimental realization of
  {E}instein-{P}odolsky-{R}osen-{B}ohm {G}edankenexperiment: {A} new violation
  of {B}ell's inequalities, Phys.\ Rev.\ Lett. \textbf{49} (1982) 91.

\bibitem{Weihs1998}
G.~Weihs, T.~Jennewein, C.~Simon, H.~Weinfurter, A.~Zeilinger, Violation of
  {B}ell's inequality under strict {E}instein locality conditions, Phys.\ Rev.\
  Lett. \textbf{81} (1998) 5039.

\bibitem{Scheidl2010}
T.~Scheidl, et~al., Violation of local realism with freedom of choice, Proc.\
  Natl.\ Acad.\ Sci.\ U.S.A. \textbf{107}~(46) (2010) 19708.

\bibitem{Giustina2013}
M.~Giustina, et~al., Bell violation using entangled photons without the
  fair-sampling assumption, Nature \textbf{497} (2013) 227.

\bibitem{Pearle1970}
P.~M. Pearle, Hidden-variable example based upon data rejection, Phys.\ Rev.\ D
  \textbf{2} (1970) 1418.

\bibitem{Garg&Mermin1987}
A.~Garg, N.~D. Mermin, Detector inefficiencies in the
  {E}instein-{P}odolsky-{R}osen experiment, Phys.\ Rev.\ D \textbf{35} (1987)
  3831.

\bibitem{Larsson1998a}
J.-{\AA}. Larsson, Bell's inequality and detector inefficiency, Phys.\ Rev.\ A
  \textbf{57} (1998) 3304.

\bibitem{Christensen2013}
B.~G. Christensen, et~al., Detection-loophole-free test of quantum nonlocality
  and applications, Phys.\ Rev.\ Lett. \textbf{111} (2013) 130406.

\bibitem{Hensen2015}
B.~Hensen, et~al., Loophole-free {B}ell inequality violation using electron
  spins separated by 1.3 kilometres, Nature \textbf{526} (2015) 682.

\bibitem{Giustina2015}
M.~Giustina, et~al., Significant loophole-free test of {B}ell's theorem with
  entangled photons, Phys.\ Rev.\ Lett. \textbf{115} (2015) 250401.

\bibitem{Shalm2015}
L.~K. Shalm, et~al., Strong loophole-free test of local realism, Phys.\ Rev.\
  Lett. \textbf{115} (2015) 250402.

\bibitem{Liu2016}
B.-H. Liu, X.-M. Hu, J.-S. Chen, Y.-F. Huang, Y.-J. Han, C.-F. Li, G.-C. Guo,
  A.~Cabello, Nonlocality from local contextuality, Phys.\ Rev.\ Lett.
  \textbf{117} (2016) 220402.

\bibitem{Cabello2010}
A.~Cabello, Proposal for revealing quantum nonlocality via local contextuality,
  Phys.\ Rev.\ Lett. \textbf{104} (2010) 220401.

\bibitem{LaCour2014}
B.~R. {La Cour}, A locally deterministic, detector-based model of quantum
  measurement, Found.\ Phys. \textbf{44} (2014) 1059.

\bibitem{LaCour2016}
B.~R. {La Cour}, A local hidden-variable model for experimental tests of the
  {GHZ} puzzle, Quantum Studies: Math.\ Found. \textbf{3} (2016) 221.

\bibitem{Mermin1990}
N.~D. Mermin, Simple unified form for the major no-hidden-variables theorems,
  Phys.\ Rev.\ Lett. \textbf{65} (1990) 3373.

\bibitem{Peres1990}
A.~Peres, Incompatible results of quantum measurements, Phys.\ Lett.\ A
  \textbf{151}~(3--4) (1990) 107.

\bibitem{EmergingQuantum}
L.~{de la Pe\~na}, A.~M. Cetto, A.~V. Hern\'andez, The Emerging Quantum: The
  Physics Behind Quantum Mechanics, Springer, Dordrecht, 2015.

\bibitem{SuppMat}
See Supplemental Material at
  \url{http://dx.doi.org/10.1016/j.physleta.2017.05.010} for a copy of the
  Matlab code used to generate the simulated data. The code may also be run
  under Octave.

\end{thebibliography}

\end{document}